\documentclass[aps,prl,twocolumn,floats,showpacs]{revtex4}

\usepackage{bm}
\usepackage{graphicx}

\def \WHOZA{Institute of Experimental Physics, Faculty of Physics,
University of Warsaw, Pasteura 5, 02-093 Warsaw, Poland}

\newcommand{\spd}{\textit{s},\textit{p}--\textit{d}\ }

\begin{document}

\title{Magnetic Ground State of an Individual Fe$^{2+}$ Ion in Strained Semiconductor Nanostructure}
\author{T.
\surname{Smole\'nski}}\email{Tomasz.Smolenski@fuw.edu.pl}\affiliation{\WHOZA}
\author{T. \surname{Kazimierczuk}}\email{Tomasz.Kazimierczuk@fuw.edu.pl}\affiliation{\WHOZA}
\author{J. \surname{Kobak}}\affiliation{\WHOZA}
\author{M. \surname{Goryca}}\affiliation{\WHOZA}
\author{A. \surname{Golnik}}\affiliation{\WHOZA}
\author{P. \surname{Kossacki}}\affiliation{\WHOZA}
\author{W. \surname{Pacuski}}\email{Wojciech.Pacuski@fuw.edu.pl}\affiliation{\WHOZA}

\date{\today}

\begin{abstract}
We investigate spin properties of a Fe$^{2+}$ dopant, known for having single nondegenerate ground state in bulk host semiconductor. Due to zero magnetic moment such a ground state is of little use for spintronics and solotronics. We show that this well-established picture of  Fe$^{2+}$ spin configuration can be contradicted by subjecting the Fe$^{2+}$ ion to sufficiently high strain, e.g., resulting from lattice mismatched epitaxial heterostructures. Our analysis reveals that high strain induces qualitative change in the ion energy spectrum and results in doubly degenerate ground state with spin projection $S_z=\pm 2$. An experimental proof of this concept is demonstrated using a new system: an epitaxial quantum dot containing individual Fe$^{2+}$ ion. Magnetic character of the Fe$^{2+}$ ground state in a CdSe/ZnSe dot is revealed in photoluminescence experiments by exploiting a coupling between a confined exciton and the single iron impurity.
\end{abstract}


\maketitle

Spin configurations of transition metal ions in various host semiconductors have been well established already a few decades ago \cite{Wieringen_DFS_1955,Low_1960_PR_theory, Low_1960_PR_experiment, Weakliem_1962_JCP, Slack_PR_1966, Baranowski_PR_1967, Gaj_SSC_1978, Furdyna_1982_JAP}. It has been found that ions such as Cr$^{2+}$($d^4$), Mn$^{2+}$($d^5$), Co$^{2+}$($d^7$) exhibit nonzero spin in their ground states, which makes them useful in spintronics \cite{Awschalom_S_2013,Dietl_RMP_2014}. However, the  ground state of the Fe$^{2+}$($d^6$) ion in zinc-blende or wurtzite II-VI semiconductors like ZnS, ZnSe, CdTe or CdSe has been found to be nondegenerate  \cite{Slack_PR_1967, Slack_PR_Fe_1969, Vallin_PRB_Fe_1970, Mycielski_1988_JAP, Scalbert_PRL_89, Scalbert_SSC_1989, Twardowski_1990_JAP, Udo_PRB_1992, Malguth_PSS_2007} and thus termed nonmagnetic  \cite{Heiman_1988_PRL}. To induce Fe$^{2+}$ magnetic moment, high magnetic field has to be applied, as for Van Vleck paramagnets \cite{Van-Vleck_1932,Mahoney_1970_JCP}.

The physics of the transition metal ions has been recently brought back into the spotlight due to possibility to access to the properties of single dopants \cite{Besombes_PRL_2004, Kudelski_2007_PRL, Flatte_NM_2011, Bocquel_PRB_2013,Yin_N_2013, Kobak_2014,Pacuski_CGD_2014}. Among other achievements, optical orientation \cite{LeGall_PRL_2009, Goryca_PRL_2009, Goryca_2010_Phys_E,Smolenski_PRB_2015}, readout \cite{Besombes_PRL_2004, Kudelski_2007_PRL, Kobak_2014, Koperski_PRB_2014} and coherent precession \cite{Goryca_PRL_2014} of a~single magnetic ion spin were demonstrated. Current development of the field benefits greatly from the fundaments of the early findings. However, the physics of the transition metal ions in semiconductor nanostructures goes far beyond the limits established in the earlier works on bulk materials.

In this Letter we demonstrate that, contrary to the well-established knowledge on a Fe$^{2+}$ ion in the semiconductor matrix, it is possible to qualitatively change its low-field behavior from nonmagnetic to magnetic, in particular by placing such an ion in a highly strained nanostructure. In order to elucidate this fact, we analyze the Fe$^{2+}$ energy spectra for the cases of weak and strong strain, showing a clear hierarchy of the energy scales, relevant both to zinc-blende and wurtzite structures. The magnetic behavior of the Fe$^{2+}$ ion is experimentally evidenced by analyzing the magnetic field dependence of the photoluminescence (PL) spectrum of an individual CdSe/ZnSe quantum dot (QD) containing  a single Fe$^{2+}$ impurity. The nonzero spin ground state of the Fe$^{2+}$ ion opens the possibility of using it as a two-level system in quantum information technology \cite{Awschalom_S_2013}.

\begin{figure}
\includegraphics{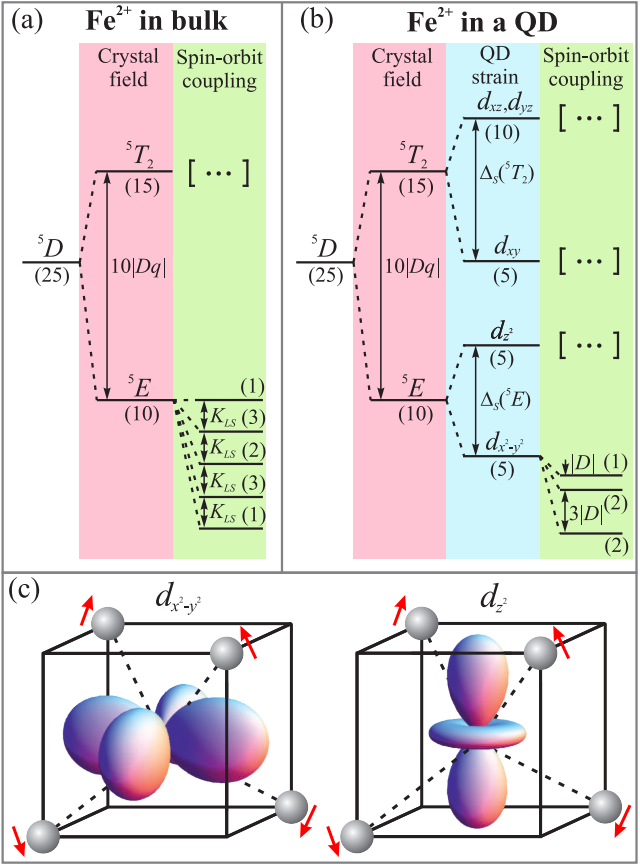}
\caption{(Color online) Energy spectrum of a Fe$^{2+}$ ion in (a) a bulk zinc-blende semiconductor, (b) a nanostructure with a strong in-plane compressive strain. Numbers in parentheses denote the degeneracy of the energy levels. Labels of the orbital states split by the QD strain refer to single-electron orbitals of corresponding symmetry (for more details, see Supplemental Material). (c) Visualization of two orbitals forming the $^5E$ subspace. Arrows schematically mark the shift of the neighboring anions due to the strain of the QD.
 \label{fig1}}
\end{figure}

The dominant effect defining energy spectrum of a transition metal ion in the bulk semiconductor is the crystal field \cite{Low_1960_PR_theory, Twardowski_1990_JAP}. Fe$^{2+}$ has configuration $d^6$, which means that the \textit{d}-shell electrons have combined orbital angular momentum of $L=2$ and spin of $S=2$. The crystal field affects only orbital part of the wave function and splits five orbital states of the ion into two subspaces: twofold degenerate $^5E$ and threefold degenerate $^5T_2$, with $^5E$ being lower-energy in $T_d$ symmetry (Fig. \ref{fig1}(a)). For Fe$^{2+}$ in CdSe or ZnSe this splitting is about $10|Dq| \approx 0.3$ eV \cite{Baranowski_PR_1967, Buhmann_PRB_1981, Malguth_PSS_2007}. Thus, the $^5T_2$ level is not populated even at room temperature and the properties of the Fe$^{2+}$ ion are defined only by the states in the $^5E$ subspace. These states are not affected by a static Jahn-Teller distortion, as it was shown for many Fe-doped semiconductors \cite{Slack_PR_1966, Slack_PR_1967, Slack_PR_Fe_1969, Vallin_PRB_Fe_1970}. Consequently, the second effect in order of strength is the spin-orbit interaction $\lambda\vec{L}\vec{S}$. It results in splitting of $^5E$ levels into 5 equidistant groups, as shown in Fig. \ref{fig1}(a). The value of the splitting is given by the effective strength of $\lambda\vec{L}\vec{S}$ interaction and the crystal field splitting: $K_\mathrm{LS} = 6\lambda^2/10|Dq| \approx 2$~meV \cite{Scalbert_SSC_1989, Udo_PRB_1992}. The presence of a dynamical Jahn-Teller effect or application of additional stress in experimentally accessible range results in only small shifts of those energy levels and can be treated perturbatively \cite{Vallin_PRB_Cr_1974, Lebecki_SSC_1991, Twardowski_SSC_1992, Udo_PRB_1992, Trushkin_NJP_2008, Malguth_PSS_2007}. In every case, the lowest-energy group consists of a single nondegenerate state, which determines the nonmagnetic character of the Fe$^{2+}$ ion ground state.

We find that strong structural strain of a QD changes hierarchy of the Fe$^{2+}$ energy scales. The dominant effect is still the crystal field, but the second effect becomes the biaxial strain. It lifts orbital degeneracy of the $^5E$ subspace, splitting it into states of symmetries corresponding to single-electron $d_{x^2-y^2}$ and $d_{z^2}$ orbitals (Fig. \ref{fig1}(b)). The ordering of those states is determined by the sign of the strain, which, given the CdSe/ZnSe lattice mismatch, has compressive character. Qualitatively, such strain pulls the tetrahedral lattice bonds away from $xy$ plane and thus lowers the energy of the $d_{x^2-y^2}$ orbital while increasing the energy of the $d_{z^2}$ one (as schematically depicted in Fig. \ref{fig1}(c)). More strict analysis leading to the same level ordering is presented in Supplemental Material (SM). Finally, the spin part of the wave function is determined by the $\lambda\vec{L}\vec{S}$ interaction. This interaction contributes to the energy of spin states of $d_{x^2-y^2}$ orbital in the second order. It favors high spin states according to the effective spin Hamiltonian $DS_z^2$ with $D < 0$ (for details, see SM). As a result, the ground state is doubly degenerate (within the discussed second order of the $\lambda\vec{L}\vec{S}$ interaction) with the spin part of $S_z = \pm 2$. Such two states are easily split by external magnetic field, in a clear contrast to previously described case of the Fe$^{2+}$ embedded in bulk semiconductor.

The experiment evidencing actual behavior of the Fe$^{2+}$ ion in a strained nanostructure is carried out on a number ($>30$) of single QDs, each incorporating an individual iron ion. Such structures are presented here for the first time. Self-assembled zinc-blende CdSe QDs in ZnSe barrier are grown using molecular beam epitaxy. About 2 monolayers of CdSe:Fe are deposited without any growth interruptions on 1.5~$\mu$m thick ZnSe buffer grown on GaAs (100) substrate. QDs are covered by a 50 nm thick ZnSe cap layer. Iron doping density is adjusted in order to optimize the probability of finding a QD with exactly one Fe$^{2+}$ ion. Low temperature ($\sim$1.5~K) PL experiments on individual QDs are performed in a setup providing spatial resolution of 0.5$~\mu$m without the need for mesas or masks. The PL is excited nonresonantly at 405~nm. The magnetic field up to 10~T is applied in Faraday configuration.

\begin{figure}
\includegraphics{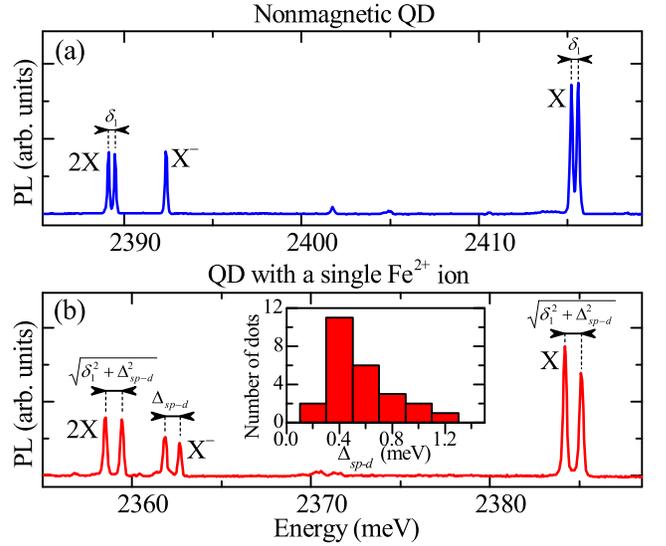}
\caption{(Color online)
(a) A PL spectrum of a typical CdSe QD showing neutral exciton (X), charged exciton (X$^-$), and biexciton (2X) lines. Neutral complexes exhibit anisotropic splitting of $\delta_1=370$ $\mu$eV. (b) A PL spectrum of a QD with a single Fe$^{2+}$ ion. The PL lines are split mainly due to \spd exchange interaction between confined carriers and the \textit{d}-shell electrons of the ion. For both spectra continuous background was subtracted. Inset: histogram of the \spd exchange splitting of the X$^-$ emission line. The cut-off at $\Delta_{s,p-d}\lesssim0.3$~meV is due to our selection procedure related to the resolution of our experimental setup --- only dots with larger zero-field splitting were tested in the magnetic field to verify the presence of the Fe$^{2+}$ ion. \label{fig:spectra}}
\end{figure}

As expected for random character of low density doping, in the same sample we find QDs incorporating single Fe$^{2+}$ ions and undoped QDs for reference purposes. PL spectra corresponding to both of these cases are shown in Fig. \ref{fig:spectra}. Figure \ref{fig:spectra}(a) presents a~typical spectrum of an undoped QD. The spectrum exhibits all standard features of epitaxial quantum dots \cite{Bayer_PRB_2002, Kulakovskii_PRL_1999, Patton_PRB_2003, Kobak_2014}. The sharp emission lines originate from recombination of different excitonic complexes, including neutral exciton (X), negatively charged exciton (X$^-$), and biexciton (2X). The neutral exciton and biexciton lines are split due to anisotropic part of electron-hole exchange interaction. In the case of QD shown in Fig. \ref{fig:spectra}(a) this splitting yields $\delta_1=370$ $\mu$eV, which is a representative value for our samples. On the other hand, the charged exciton line does not exhibit any splitting, in accordance with the Krammers rule for systems with odd number of fermions.

In comparison, introduction of a single Fe$^{2+}$ ion into a QD leads to distinctive changes in the PL spectrum, as shown in Fig. \ref{fig:spectra}(b). The emission lines still correspond to recombination of the same excitonic complexes, however their structure is determined by the \spd exchange interaction with the resident ion. The main effect is a strong splitting of each of the observed emission lines. It is particularly striking for typically degenerate charged exciton, but also for the neutral exciton it is significantly stronger than typical value of $\delta_1$. Such a physical picture is similar for large number of studied Fe-doped QDs, as proven by distribution of measured \spd exchange splittings presented in the inset to Fig. \ref{fig:spectra}(b). The presence of such \spd splitting is a direct confirmation of the magnetic character of the Fe$^{2+}$ ion. It originates from the fact that the Fe$^{2+}$ spin may be aligned either parallel or anti-parallel to the exciton angular momentum, which would not be possible in the case of nonmagnetic ground state.

\begin{figure}
\includegraphics{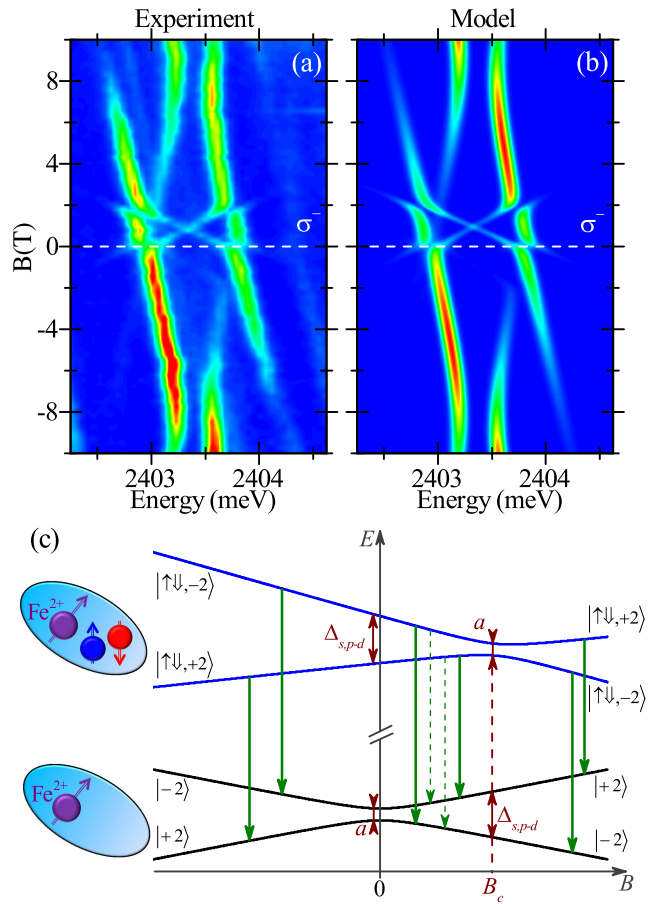}
\caption{(Color online) Magnetic field dependence of the PL spectrum of a neutral exciton in a QD with a single Fe$^{2+}$ ion: (a) experimental data, and (b) simulation assuming strain-induced magnetism of the Fe$^{2+}$ ion, as described in the text. The spectra were measured (simulated) in $\sigma^{-}$ circular polarization. (c) Schematic dependence of the involved energy levels on the magnetic field together with indicated $\sigma^-$ polarized optical transitions observed in PL measurements. The upper pair of levels corresponds to $\left|\uparrow \Downarrow\right\rangle$ exciton coupled with the ion spin (where $\uparrow$ and $\Downarrow$ represent the spin projection of the electron and the heavy hole on the growth axis, respectively), while the bottom pair represents the energies of the ion states in the empty dot. The excitonic transitions preserving (altering) the ion spin projection are marked with solid (dashed) arrows.\label{fig:pole}}
\end{figure}

In order to provide the final proof of the magnetic character of the Fe$^{2+}$ ion in a QD, we measure the evolution of the X PL spectrum in external magnetic field applied along the growth direction (quantization axis of the magnetic ion and QD excitons). Typical results obtained in $\sigma^-$ polarization of detection are shown in Fig. \ref{fig:pole}(a). We note that the observed pattern is quite similar to the one obtained for InAs/GaAs QDs containing single manganese ions \cite{Kudelski_2007_PRL,Krebs_PRB_09}, despite different microscopic origin.

Magneto-photoluminescence results in Fig. \ref{fig:pole}(a) seem complex, however they originate from quite simple behavior of the initial and final energy levels of the transitions, as illustrated in Fig. \ref{fig:pole}(c). First effect of the magnetic field is the Zeeman splitting between $S_z = 2$ and $S_z = -2$ states of the Fe$^{2+}$ ion. Unfortunately, the PL spectrum does not show this splitting directly, since in general exciton recombination does not influence the ion spin state and thus the energy of emitted photon  does not depend on the ion Zeeman splitting. However, in the vicinity of the level anticrossings the Fe$^{2+}$ spin states are mixed and this selection rule is relaxed. Indeed, data in Fig. \ref{fig:pole}(a) features several weaker lines in the anticrossing range (i.e., 0 -- 2~T). Before we discuss the origin of the anticrossings, let us analyze the behavior of these weak lines, in particular the cross-like feature. The two crossing lines correspond to transitions involving the change of the ion spin from $S_z = \pm 2$ to $S_z=\mp 2$. The splitting between them depends almost linearly on the magnetic field with a slope of about $0.84$~meV/T. More precise fitting including non-linearity due to proximity of the anticrossings gives slightly larger value of $0.92$~meV/T. Taking into account that $\left|\Delta S_z\right| = 4$ for both the initial and the final states, this slope corresponds to g-factor of 2.0, exactly as expected for the Fe$^{2+}$ spin.

Let us now focus on the nature of the observed anticrossings. The first, relatively weak anticrossing occurs around 0~T. It is a signature that the $S_z=\pm 2$ states of the Fe$^{2+}$ ion are not perfectly degenerate, but are split by a small energy $a$, as shown in Fig. \ref{fig:pole}(c). This splitting varies between different studied dots and its average value yields about 50~$\mu$eV. Consequently, the presence of this splitting does not invalidate our conclusion about magnetism of the ground state of the Fe$^{2+}$ ion, since $a$ is negligible even compared with the X--Fe$^{2+}$ exchange. Such a~splitting requires including $\lambda\vec{L}\vec{S}$ coupling in the fourth order of perturbation theory, according to which the zero-field eigenstates are $\frac{1}{\sqrt{2}} \left( \left|S_z=2 \right> \pm \left|S_z=-2 \right> \right)$.

The second anticrossing around 2~T is closely related to the first one. It occurs when the effective magnetic field acting on the Fe$^{2+}$ spin in the presence of the $\sigma^-$ emitting exciton vanishes. Since exchange field of this exciton increases the energy of the state corresponding to $S_z=-2$ ion spin projection (Fig. \ref{fig:pole}(c)), the anticrossing of the Fe$^{2+}$ ion is effectively shifted from 0~T to a higher field.

Finally, there is also the third, stronger anticrossing around $\pm$9~T. This anticrossing is observed for both negative and positive magnetic field (or equivalently: in $\sigma^+$ and $\sigma^-$ polarization), which clearly indicates that it is due to mixing of the exciton part of the total wave function. Indeed, the states involved in the anticrossing correspond to $\sigma^-$ and $\sigma^+$ emitting excitons coupled with $S_z=-2$ spin projection of the Fe$^{2+}$ ion ($\left| \uparrow \Downarrow,-2 \right>$ and $\left| \downarrow \Uparrow,-2 \right>$). The anticrossing occurs when the excitonic Zeeman effect reduces the ion-related exchange splitting of the involved states and the anisotropic electron-hole exchange interaction becomes dominant source of the splitting. It should be noted that this anticrossing does not mix different states of the Fe$^{2+}$ ion and thus in this range of magnetic field the optical recombination preserves the spin of the ion.

In order to quantitatively verify our interpretation of the magneto-photoluminescence results, we perform a numerical simulation of the expected field-dependence of X PL spectrum. The simulation is based on the general spin Hamiltonian of an ion-exciton system \cite{Besombes_PRL_2004, Kudelski_2007_PRL,Kobak_2014}:
\begin{displaymath}
H = H_\mathrm{ion} + H_\mathrm{X} + H_{s,p-d} + \mu_B B \left( g_\mathrm{ion} S^\mathrm{ion}_z + g_e S^\mathrm{e}_z + g_h S^\mathrm{h}_z \right),
\end{displaymath}
where $H_\mathrm{ion}$ is the Hamiltonian of the Fe$^{2+}$ ion leading to energy spectrum as in Fig \ref{fig1}(b), $H_\mathrm{X}$ is the Hamiltonian of the exciton with electron-hole exchange interaction \cite{Bayer_PRB_2002}, $H_{s,p-d}$ describes the \spd exchange interaction between confined carriers and the iron ion \cite{Twar90, Scal90, Test91, Test00, Pacuski_PRL08}, and $g_\mathrm{ion}$, $g_e$ and $g_h$ are g-factors of the Fe$^{2+}$ ion, electron and hole, respectively. As shown in Fig. \ref{fig:pole}(b), such a model reproduces all features of the experimental measurement (for details of the simulation procedure, see SM). The model correctly captures even the observed thermalisation of the ion spin at increasing magnetic field by taking into account the effective Fe$^{2+}$ spin temperature of 15~K. Such a good overall agreement provides a strong proof of correct identification of all relevant effects.

All the presented results clearly show that the structural strain of the QD induces magnetic character of the Fe$^{2+}$ ion in its ground state. Our findings reveal that a CdSe/ZnSe QD containing such an ion may be useful for spin-based quantum information technology, as it combines many desired qualities. In particular, both the CdSe lattice and Fe$^{2+}$ ion can be free of any nuclear spin fluctuations. Moreover, the QD provides efficient optical access to a~single ion. Finally, as we show here, the ground state of the Fe$^{2+}$ ion has nonzero spin, which opens the possibility of using such ion as a two-level system. However, the importance of our results is not limited to this particular system. It is a~general example of the fact that even well-established textbook knowledge of energy spectrum of various dopants should be carefully re-evaluated in the world of semiconductor nanostructures.

\begin{acknowledgments}
We thank A. Twardowski and T. Dietl for helpful discussions as well as M. Dobrowolski and M. Koperski for experimental assistance in the first measurements. This work was supported by the Polish National Science Center under decisions DEC-2011/01/B/ST3/02406, DEC-2011/02/A/ST3/00131, DEC-2013/09/B/ST3/02603, DEC-2013/09/D/ST3/03768, DEC-2012/05/N/ST3/03209 and DEC-2012/07/N/ST3/03130, by the Polish Ministry of Science and Higher Education in years 2012-2016 as research grant "Diamentowy Grant", by the Polish National Centre for Research and
Development project LIDER, and by the Foundation for Polish Science (MISTRZ programme). Project was carried out with the use of CePT, CeZaMat, and NLTK infrastructures financed by the European Union - the European Regional Development Fund within the Operational Programme "Innovative economy" for 2007 - 2013.
\end{acknowledgments}

\end{document}

\title{Supplemental Material for ``Magnetic Ground State of an Individual Fe$^{2+}$ Ion in Strained Semiconductor Nanostructure''}
\author{T.
\surname{Smole\'nski}}\email{Tomasz.Smolenski@fuw.edu.pl}\affiliation{\WHOZA}
\author{T. \surname{Kazimierczuk}}\email{Tomasz.Kazimierczuk@fuw.edu.pl}\affiliation{\WHOZA}
\author{J. \surname{Kobak}}\affiliation{\WHOZA}
\author{M. \surname{Goryca}}\affiliation{\WHOZA}
\author{A. \surname{Golnik}}\affiliation{\WHOZA}
\author{P. \surname{Kossacki}}\affiliation{\WHOZA}
\author{W. \surname{Pacuski}}\email{Wojciech.Pacuski@fuw.edu.pl}\affiliation{\WHOZA}

%\date{\today}

%\pacs{78.67.Hc, 78.55.Et, 75.75.-c}

\maketitle

\section{Labeling of Fe$^{2+}$ ion orbital states}

In Table \ref{tab1} we list the symmetries of five multi-electron orbital states of Fe$^{2+}$ ion. Since six $d$-shell electrons have total orbital angular momentum $L=2$, their orbital wave functions transform like linear combinations of spherical harmonics $Y_2^{L_z}$, where $L_z=0,\ \pm1,\ \pm2$. They are explicitly given in the second column of Table \ref{tab1} and denoted by symbols $\eta$, $\xi$, $\zeta$, $\theta$, $\epsilon$ after J. T. Vallin \textit{et al.} \cite{Vallin_PRB_Fe_1970}. Since $d^6$ configuration corresponds to half-filled shell with one additional electron, these multi-electron orbital states may be also discussed using single-electron $d$ orbitals of the same symmetries, which are listed in the last column.

\begin{table}[h]
\begin{center}
\begin{tabular}{|c|c|c|}
\hline
\multirow{3}{*}{Subspace} & \multirow{3}{*}{Orbital wave function} & \multirow{3}{*}{\parbox{2.2cm}{Corresponding single-electron orbital}} \\
& & \\
& & \\\hline\hline
\multirow{3}{*}{$^5T_2$} & $|\eta\rangle=-(Y_2^{1}-Y_2^{-1})/\sqrt{2}$ & $d_{xz}$ \\\cline{2-3}
& $|\xi\rangle=i(Y_2^{1}+Y_2^{-1})/\sqrt{2}$ & $d_{yz}$ \\\cline{2-3}
& $|\zeta\rangle=i(Y_2^{2}-Y_2^{-2})/\sqrt{2}$ & $d_{xy}$ \\\hline\hline
\multirow{2}{*}{$^5E$} & $|\theta\rangle=Y_2^0$ & $d_{z^2}$ \\\cline{2-3}
& $|\epsilon\rangle=(Y_2^{2}+Y_2^{-2})/\sqrt{2}$ & $d_{x^2-y^2}$\\\hline
\end{tabular}
\end{center}
\caption{Symmetry of Fe$^{2+}$ orbital states. \label{tab1}}
\end{table}

\section{Energy levels of Fe$^{2+}$ ion in the presence of biaxial strain}
\begin{figure}
\includegraphics{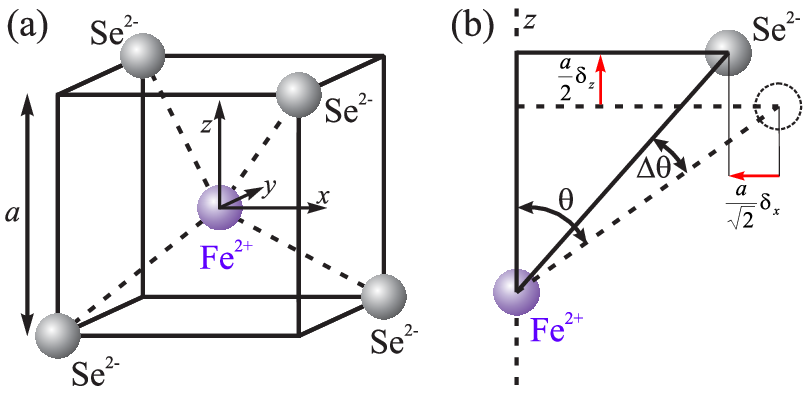}
\caption{(a) A schematic visualisation of Fe$^{2+}$ ion surrounded tetrahedrally by four Se$^{2-}$ anions occupying alternating corners of a cubic elementary cell. (b) Illustration of strain-induced displacement of each Se$^{2-}$ anion (not in scale). The angle $\theta$ corresponds to the anion position in pure $T_d$ symmetry, while $\Delta\theta$ denotes the change arising due to the strain.
 \label{fig1}}
\end{figure}

In order to estimate the magnitude of strain-induced splitting of $^5E$ orbitals we employ a simple point-charge model, in which the crystal field acting on the single Fe$^{2+}$ ion comes from four point charges corresponding to neighboring Se$^{2-}$ anions (Fig. \ref{fig1}(a)). The electrostatic potential from these charges can be expanded in multipole series. In the case of pure tetrahedral $T_d$ symmetry, the terms relevant to $d$ orbitals of the Fe$^{2+}$ ion are given by \cite{Low_1960_PR_theory, Slack_PR_Fe_1969, Vallin_PRB_Cr_1970}
\begin{equation}
V_C(\vec{r})=A\frac{r^4}{d^4}\left[Y_4^0+\sqrt{\frac{5}{14}}\left(Y_4^{4}+Y_4^{-4}\right)\right],
\end{equation}
where $Y_n^m(\theta, \varphi)$ are spherical harmonics in a coordinate system $(r,\theta,\varphi)$ centered on the Fe$^{2+}$ ion. The coefficient $A$ denotes the strength of the crystal field and is given by $A=-\frac{56\sqrt{\pi}}{27}\frac{e^2}{4\pi\epsilon_0\epsilon_rd}$, where $\epsilon_r$ is the relative permittivity of CdSe taken as 9.2 \cite{Geick_JAP_1966} and $d=a\sqrt{3}/2$ is the distance between Fe$^{2+}$ and each anion in the crystal lattice (with $a\approx6.08$~{\AA} being zinc-blende CdSe lattice constant \cite{Samarth_APL_1989}).

The structural strain of a CdSe/ZnSe QD shifts the anions away from their normal positions and thus lifts the tetrahedral symmetry. The main factor determining the character of the strain is a large lattice mismatch between CdSe and ZnSe, which is equal to about 7\% \cite{Samarth_APL_1989, ACSA_McIntyre_1980}. As a result, the dot material is compressed in $xy$ plane and stretched along the growth direction $z$. We assume that the relative reduction of a lattice constant is the same for $x$ and $y$ directions and equal to $\delta_x$. Consequently, the relative elongation of an elementary cell in $z$ is given by $\delta_z=2\nu \delta_x$, where $\nu\approx0.35$ is CdSe Poisson's ratio \cite{Deligoz_PhysB_2006}. Under these assumptions, the angular coordinate $\varphi$ of each anion in $xy$ plane remains unchanged, while the angle $\theta$ between the $z$ axis and direction of Fe$^{2+}$--Se$^{2-}$ bond is slightly decreased, as schematically shown in Fig. \ref{fig1}(b). To the accuracy of linear terms in $\delta_x$, the change of $\theta$ angle is given by $\Delta \theta=\sqrt{2}(1+2\nu)\delta_x/3$, which for maximal $\delta_x$ limited by the lattice mismatch corresponds to about 3$^\circ$.

Such strain-induced displacement of the anions lowers the symmetry to $D_{2d}$ and introduces new terms to the crystal field potential, which for $d$ orbitals read
\begin{multline}
V_S(\vec{r})=A_\theta\frac{r^2}{d^2}\cdot\\\cdot\left[Y_2^0-\frac{5\sqrt{5}}{27}\frac{r^2}{d^2}
\left(Y_4^0-\sqrt{\frac{7}{10}}\left(Y_4^4+Y_4^{-4}\right)\right)\right],
\end{multline}
where $A_\theta\approx16\sqrt{\frac{2\pi}{5}}\frac{e^2}{4\pi\epsilon_0\epsilon_rd}\Delta\theta$. The overall potential $V_C+V_S$ splits fivefold degenerate orbitals of the Fe$^{2+}$ ion into four subspaces, which energies take a form
\begin{eqnarray}
E(d_{xz})=E(d_{yz})&=&-4Dq+Ds+2Dt,\label{dxz} \\
E(d_{xy})&=&-4Dq-2Ds-4Dt,\label{dxy} \\
E(d_{z^2})&=&6Dq+2Ds-3Dt, \label{dz2} \\
E(d_{x^2-y^2})&=&6Dq-2Ds+3Dt, \label{dx2y2}
\end{eqnarray}
where $10|Dq|$ (with $Dq<0$) is a splitting between $^5T_2$ and $^5E$ subspaces arising only due to the crystal field of pure $T_d$ symmetry, while $Ds$ and $Dt$ are the splittings induced by the strain. They are given by
\begin{eqnarray}
Dq&=&-\frac{4}{27}\frac{e^2}{4\pi\epsilon_0\epsilon_rd^5}\langle r^4\rangle,\label{Dq}\\
Ds&=&\frac{16}{21}(1+2\nu)\delta_x\frac{e^2}{4\pi\epsilon_0\epsilon_rd^3}\langle r^2\rangle,\label{Ds}\\
Dt&=&\frac{160}{567}(1+2\nu)\delta_x\frac{e^2}{4\pi\epsilon_0\epsilon_rd^5}\langle r^4\rangle,\label{Dt}
\end{eqnarray}
where $\langle r^n\rangle$ denotes the expectation value of $n$th power of Fe$^{2+}$ orbital radius $r$. According to Eqs. (\ref{dz2}) and (\ref{dx2y2}), the strain-induced energy splitting of orbitals $d_{x^2-y^2}$ and $d_{z^2}$ forming the $^5E$ subspace corresponds to ${\Delta_S(^5E)=4Ds-6Dt}$. Under an assumption that $\langle r^n\rangle\approx \langle r\rangle^n$ it can be expressed using $|Dq|$ and reads
\begin{multline}
\Delta_S(^5E)=(1+2\nu)\delta_x\frac{80}{7}|Dq|\cdot\\\cdot\left(\frac{2\sqrt{3}}{5}\sqrt{\frac{e^2}{4\pi\epsilon_0\epsilon_rd|Dq|}}-1\right).\label{delta5E}
\end{multline}
Since $10|Dq|\approx0.3$~eV for CdSe \cite{Baranowski_PR_1967, Buhmann_PRB_1981, Malguth_PSS_2007}, the splitting $\Delta_S(^5E)$ corresponds to $\delta_x\cdot0.7$~eV, which for maximal $\delta_x$ of 0.07 yields about 50~meV. This value is significantly larger than the splitting of $^5E$ orbitals arising from spin-orbit coupling with higher energy states, which was found to be of the order of a few meV in bulk semiconductors \cite{Scalbert_SSC_1989, Udo_PRB_1992, Malguth_PSS_2007}. Consequently, our calculations reveal that the crystal field and strain are dominant effects influencing the energies of the Fe$^{2+}$ ion states in a CdSe/ZnSe QD and the spin-orbit interaction might be treated perturbatively. Moreover, our results also confirm that $d_{x^2-y^2}$ is the lowest-energy orbital state of this ion embedded in a dot, since the computed $\Delta_S(^5E)$ is positive for the compressive strain (i.e., for $\delta_x>0$). It should be noted that the actual value of this splitting may be deviated to a some degree from the one given by Eq. (\ref{delta5E}), since the exploited point-charge model neglects the covalency of bonds between magnetic ion and anions in the crystal lattice. Nevertheless, as proven by the experiment, the general conclusions of our analysis are not altered by this simplification, even though it is probably quite important, especially in the case of II-VI compounds studied in this work.

\section{Effects of spin-orbit coupling}

Since both the crystal field and strain does not affect the spin part of the Fe$^{2+}$ ion states, the lowest-energy orbital  $d_{x^2-y^2}$ is fivefold degenerate due to the ion spin $S=2$. This degeneracy is partially lifted by the spin-orbit interaction $\lambda \vec{L}\vec{S}$. Such interaction splits the spin states of $d_{x^2-y^2}$ orbital in the second order through mixing with higher energy orbitals forming the $^5T_2$ subspace. This splitting can be expressed by the effective spin Hamiltonian:
\begin{equation}
H_{LS}=D\left[S_z^2-\frac{1}{3}S(S+1)\right],
\end{equation}
where $D$ is given by
\begin{equation}
D=\frac{\lambda^2}{E(d_{xz})-E(d_{x^2-y^2})}-\frac{4\lambda^2}{E(d_{xy})-E(d_{x^2-y^2})}.
\end{equation}
Taking into account Eqs. (\ref{dxz})-(\ref{dx2y2}) and bearing in mind that $Ds$ and $Dt$ are positive for the compressive strain, one obtains $E(d_{xz})>E(d_{xy})>E(d_{x^2-y^2})$, which yields $D<-3\lambda^2/[E(d_{xy})-E(d_{x^2-y^2})]<0$. Consequently, the states corresponding to $S_z=\pm2$ spin projections have lowest energy. Their degeneracy is however lifted, when the spin-orbit coupling is considered in the higher order. In particular, in the fourth order the states ${|S_z=\pm2\rangle}$ are split by $\lambda \vec{L}\vec{S}$ interaction into linear combinations $\frac{1}{\sqrt{2}}(|S_z=2\rangle\pm|S_z=-2\rangle)$, with the state corresponding to ``+'' being lower energy. The corresponding energy splitting $a$ is given by
\begin{equation}
a=\frac{36\lambda^4}{\left[E(d_{xz})-E(d_{x^2-y^2})\right]^2\Delta_S(^5E)}.
\end{equation}
The value of this splitting obtained for maximal strain (i.e., $\delta_x=0.07$) and $\lambda\approx-10$~meV \cite{Baranowski_PR_1967, Udo_PRB_1992} is equal to about 55~$\mu$eV, which is consistent with our experimental results.

\section{Simulation of the PL spectrum of a QD with a single Fe$^{2+}$ ion}

As shown in the manuscript, all effects observed in the magnetic field evolution of the PL spectrum of a QD with Fe$^{2+}$ can be reproduced by a numerical simulation. Our simulation is based on the standard procedure of finding the eigenstates of the exciton complex and the empty (i.e., without the exciton) QD and subsequent calculation of allowed transitions. For simplicity, we assume that the spatial part of exciton wavefunction is not substantially modified by the magnetic field and we can restrict our analysis to the spin degree of freedom. The possible initial states were found by diagonalization of the Hamiltonian:
\begin{multline}
H_i = H_\mathrm{ion} + H_\mathrm{X} + H_{s,p-d} +\\+ \mu_B B \left( g_\mathrm{ion} S^\mathrm{ion}_z + g_e S^\mathrm{e}_z + g_h S^\mathrm{h}_z \right)
\end{multline}
where $S^\mathrm{ion}$, $S^\mathrm{e}$, $S^\mathrm{h}$ are effective spin operators of the ion and carriers with eigenvalues of $\pm2$, $\pm1/2$, $\pm3/2$, respectively. The Hamiltonian is diagonalized within the space of possible orientations of bright exciton and the Fe$^{2+}$ spin: $\left|\uparrow \Downarrow, +2\right\rangle$, $\left|\uparrow \Downarrow, -2\right\rangle$, $\left|\downarrow \Uparrow, +2\right\rangle$, $\left|\downarrow \Uparrow, -2\right\rangle$. Dark excitons were not included in the simulation, since due to the axial symmetry of the Hamiltonian they do not mix with the bright excitons.

The Hamiltonian of the final state includes only the contributions from the Fe$^{2+}$ ion:
\begin{equation}
H_f = H_\mathrm{ion} + g_\mathrm{ion} \mu_B B S^\mathrm{ion}_z.
\end{equation}

The Hamiltonians are parameterized by 5 free parameters: $\delta_1$, $g_X=3g_h-g_e$, $g_\mathrm{ion}$, the zero-field splitting $a$ of the Fe$^{2+}$ states, and the exciton-ion exchange splitting $\Delta_{s,p-d}$. The values of these parameters are found by fitting the field dependence of the PL spectrum. Additional parameter $E_0$ is introduced later to plot the simulation in the same energy range as the experimental results.

Since spatial part of all considered states is the same, the oscillator strength between an initial and a final state is determined only by the spin part of the wavefunction. In our case the oscillator strength is proportional to the projection of the initial state $\left|i\right\rangle$ onto the state corresponding to the final spin projection of the ion $\left|f\right\rangle$ (e.g., $\left|+2\right\rangle$) and the given orientation of the exciton spin (e.g., $\left|\uparrow\Downarrow\right\rangle$ for simulation of $\sigma-$ spectrum).

Finally, the optical spectrum is simulated as a series of peaks at energies corresponding to possible transitions between possible initial and final states. The intensity of a given peak is the product of calculated oscillator strength and the Boltzmann factor corresponding to thermalisation of the ion spin at an effective temperature of 15~K. For the sake of the presentation, the peaks are plotted as gaussians with FWHM of 0.1~meV. Strictly speaking, such procedure gives us the low-power absorption spectrum, but it is also similar to the PL spectrum if only the exciton states have the same lifetime and that is shorter than the thermalisation of the excited state.